# Title: Attosecond spectroscopy reveals alignment dependent core-hole dynamics in the ICl molecule.

## Authors:

Hugo J. B. Marroux,[1,2,†,*] Ashley P. Fidler,[1,2] Aryya Ghosh,[4] Yuki Kobayashi,[1] Kirill Gokhberg,[4] Alexander I. Kuleff,[4,5] Stephen R. Leone,[1,2,3*] Daniel M. Neumark[1,2*]

## Affiliations:

[1]Department of Chemistry, University of California, Berkeley, 94720, USA.
[2]Chemical Sciences Division, Lawrence Berkeley National Laboratory, Berkeley, 94720, USA.
[3]Department of Physics, University of California, Berkeley, 94720, USA
[4]Theoretische Chemie, PCI, Universität Heidelberg, Im Neuenheimer Feld 229, 69120, Heidelberg, Germany
[5]ELI-ALPS, W. Sandner utca 3, 6728 Szeged, Hungary
*Correspondence to: hugo.marroux@epfl.ch; srl@berkeley.edu; dneumark@berkeley.edu
†Current address: Laboratoire de Spectroscopie Ultrarapide (LSU) and Lausanne Centre for Ultrafast Science (LACUS), Ecole Polytechnique Fédérale de Lausanne, ISIC, FSB, Station 6, CH-1015 Lausanne, Switzerland

## Abstract:

The removal of electrons located in the core shells of molecules creates transient states that live between a few femtoseconds to attoseconds. Owing to these short lifetimes, time-resolved studies of these states are challenging and complex molecular dynamics driven solely by electronic correlation are difficult to observe. Here, few-femtosecond core-excited state lifetimes of iodine monochloride are obtained by attosecond transient absorption on iodine $4d^{-1}6p$ transitions around 55 eV. Core-level ligand field splitting allows direct access of excited states aligned along and perpendicular to the ICl molecular axis. Lifetimes of 3.5±0.4 fs and 4.3±0.4 fs are obtained for core-hole states parallel to the bond and 6.5±0.6 fs and 6.9±0.6 fs for perpendicular states, while nuclear motion is essentially frozen on this timescale. Theory shows that the dramatic decrease of lifetime for core-vacancies parallel to the covalent bond is a manifestation of non-local interactions with the neighboring Cl atom of ICl.

## Main text:

High energy photons in the soft x-ray regime access electrons located in core-shells of atoms and molecules. In atoms, the resulting core-excited states decay in tens of femtoseconds or less through emission of a secondary electron driven by electronic correlation.[1,2] Measurement of these excited state lifetimes in molecules through linewidth studies is challenging as lineshape analysis requires consideration of unresolved vibrational structure in addition to lifetime broadening.[3] Attosecond spectroscopy offers the possibility to measure core state lifetimes in the time domain[4] but has so far only been employed for isolated atoms[5,6] or focused on strong field related effects in core-excited molecular systems.[7] Here, we apply attosecond transient absorption spectroscopy (ATAS) to investigate the decay of core-excited states in the ICl molecule. We find that electronic decay occurs before significant nuclear motion, and that the alignment of the core-level orbital with respect to the internuclear axis has a large effect on lifetimes.

In heteronuclear molecular systems, extreme ultraviolet (XUV) or soft x-ray absorption at an element edge creates a vacancy in a specific atom, producing a charge imbalance across the molecule. The decay of the core-hole can then occur via local (i.e. atomic-like) pathways such as Auger decay, or via non-local channels involving electron or energy transfer with nearby atoms. These non-local channels lead to the emission of secondary electrons via processes such as interatomic Coulombic decay[8] or electron-transfer-mediated decay mechanisms.[9] The few time resolved experiments on non-local decay channels obtain lifetimes > 100 fs, [10–12] but those were focused on the study of inner-shell transitions of rare-gas dimers. In molecules, the small distances between the atom in which the initial excitation is created and the neighboring atoms can bring the lifetime of the core-hole down to the few or subfemtosecond regime.[3]

Ligand-field splittings of core-levels are typically a few hundreds of meV, in principle enabling the excitation of states with different alignments along the molecular symmetry axis.[13] Short lifetimes of core-hole states combined with vibrational broadening prevent the use of frequency domain techniques to accurately investigate these closely spaced states, and computational studies show that variations in lifetimes with the hole direction are expected in rare-gas dimers and are a manifestation of non-local effects.[14,15]

In time domain experiments such as ATAS, an isolated attosecond pulse (IAP) in the XUV is linearly absorbed by the sample, creating a macroscopic polarization as a result of a coherent superposition between the ground and core excited states, in this case for ICl.[16] The polarization is perturbed by a delayed near-infrared (NIR) pulse, modifying the XUV pulse absorption spectrum. By scanning the delay between the isolated attosecond pulse (IAP) and the NIR pulse, core-hole lifetimes can be retrieved from the polarization decay, as has been shown for inner valence excited states of argon and xenon[17,18] as well as core states of krypton.[6] ATAS was recently applied to core-excited states of methyl iodide, where transitions to Rydberg excited states dominate the spectrum, but no core-hole lifetimes were reported.[7]

Here we use an IAP to excite iodine $6p \leftarrow 4d$ core-to-Rydberg transitions around 55 eV to create energetically distinct states with iodine core-hole orbitals aligned parallel or perpendicular to the molecular axis for state-specific lifetime probing.[19,20] Core-hole orbitals aligned parallel to the internuclear axis exhibit significantly shorter lifetimes (3.5 ± 0.4 and 4.3 ± 0.4 fs) compared to core-hole orbitals perpendicular to the molecular axis (6.5 ± 0.6 and 6.9 ± 0.6 fs). During the timescale of these decays, the bond length changes by at most 3.5% (0.075Å), so nuclear motion is minimal. The lifetime dependence on core-hole orbital alignment is reproduced in part by ab initio calculations using the Fano-algebraic diagrammatic construction (ADC)-Stieltjes method.[21,22] Results from the calculation attribute the differences in decay rates to a greater participation of delocalized molecular orbitals (MO) (non-local effect) in the decay of core-excited states aligned along the molecular axis. The observation of excited state decays that are faster than nuclear motion and the dependence of decay rates on orbital alignment opens an uncharted field of investigation exploring electronic molecular decay dynamics using attosecond spectroscopy.

## Results and Analysis

### Static measurement

The absorption spectrum of ICl corresponding to $4d^{-1}6p$ Rydberg excitation on the I atom is collected by spectrally resolving the IAP transmitted through the sample as shown in Fig. 1a. This spectrum, spanning 54 to 59 eV, is similar to published spectra of iodine-containing diatomic molecules.[19] As shown in Fig. 1c, the iodine $4d$ core-levels are split by 1.7 eV due to spin-orbit coupling, and each spin-orbit level has a ligand-field splitting of 0.3 eV.[13]

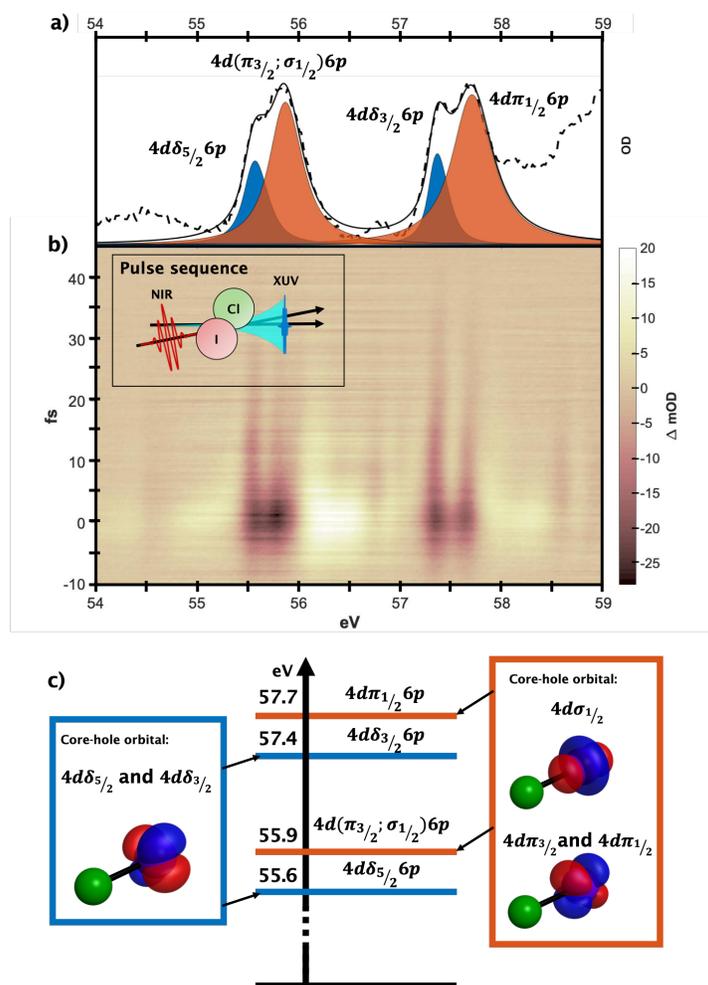

**Figure 1)**. **a)** Iodine $N_{4,5}$ edge absorption of ICl centered around $4d^{-1}6p$ transitions **b)** ATAS spectrum of ICl at the iodine $N_{4,5}$ edge. The pulse sequence of the ATAS experiment is shown in the inset where the IAP (in blue) precedes the NIR few cycle pulse (in red). The macroscopic polarization created by the IAP and perturbed by the NIR pulse is shown as a decaying feature following the IAP. **c)** Core-hole orbital wavefunctions, (cf. SM5 for computation details) separated: into orbitals perpendicular to the molecular axis in the blue box, and orbitals aligned along the molecular axis in the orange box. The potential energies are determined based on the Gaussian fittings to the experimental absorption spectrum.

**Time resolved measurement**

Fig. 1b shows the ATAS spectrum of ICl for various time delays acquired using the experimental procedure described previously[23] and in the Methods section. Here, positive time delays correspond to the XUV pulse arriving before the NIR pulse. The signal is composed of negative transient features at central frequencies of the transitions observed in the static spectrum (Fig. 1a), and positive features on either side of these.

The spectra in Fig. 1b reflect perturbation of the XUV-induced polarization by the NIR few cycle pulse.[24] Perturbation of the macroscopic polarization can proceed via laser-induced ionization of the decaying dipole,[25] resonant coupling with a dark state,[26] or non-resonant AC stark shifting of the excited state.[24] In the work reported here, single photon ionization is inaccessible with the NIR pulse and the peak power was kept low enough ($2 \times 10^{13}$ W/cm$^2$) to minimize ionization by strong field processes. The 6$s$ state, which is the principal candidate for resonant coupling with the 6$p$ excited states, is too close in energy (0.8 eV) and cannot be populated by the broadband NIR photon energy.[19] In a study of methyl iodide, Drescher et al.[7] made similar observations and came to the conclusion that a non-resonant (i.e. Stark shift) interaction is responsible for the transient signal.

In order to confirm that non-resonant coupling of the NIR pulse is the origin of the transient signal, the time-zero transient spectrum has been simulated in Fig. 2a by considering that the NIR pulse induces a shift in the phase ($\Delta\varphi$) of the macroscopic polarization that is proportional to the ponderomotive energy of the NIR field: $\Delta\varphi(t,\tau) = \int \frac{[E_0(\tau,t')]^2}{4\omega^2} dt'$ where $E_0$ and $\omega$ are the NIR field amplitude and central frequency, respectively.[24,26,27] The time-dependent Schrödinger equation is solved considering only the four main transitions from the core-levels to the 6$p$ Rydberg state. Here the features are uniformly lifetime-broadened to match the experimental spectrum and no vibronic or experimental broadening is taken into account. The main features of the experimental spectrum (in red) are reproduced in the simulation (in grey) confirming that the non-resonant interaction dominates.

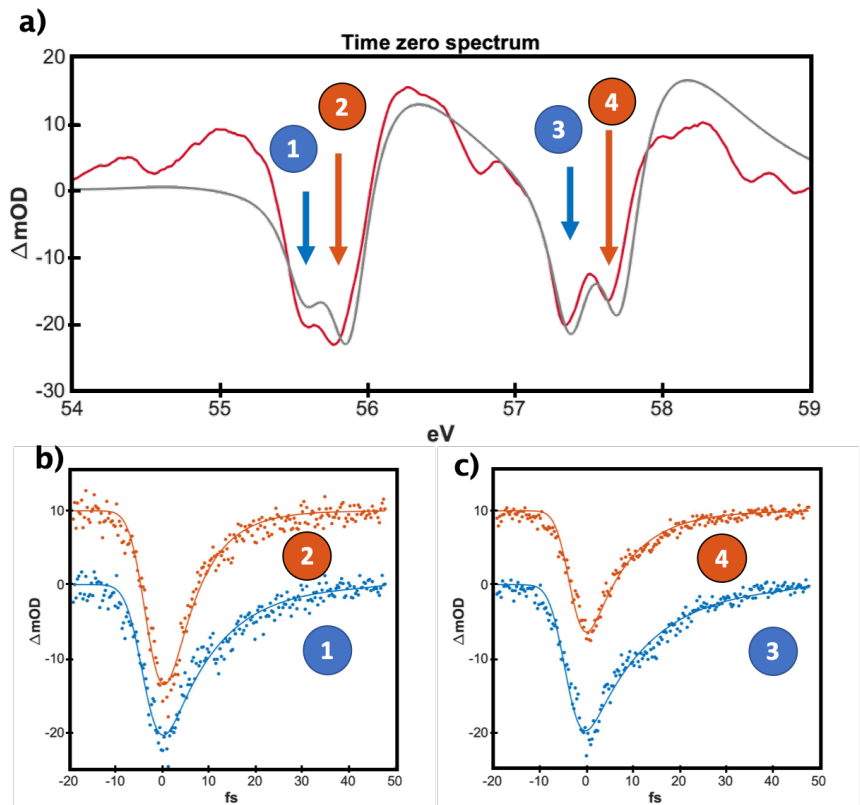

**Figure 2 a)** Time zero ATAS experimental (in red) and simulated (in grey) spectra of ICl. Kinetic traces of the depletion recovery in **b)** and **c)** at the positions indicated by the color-coded arrows and corresponding numbers in Table 1. Traces number two and four have been displaced vertically by 10 mOD for clarity.

The exponential recovery of the negative depletion features depends on the dephasing time of the considered transition.[18] In the gas phase, dephasing is limited by the population lifetime ($T_1$) of the excited states so that the decay of the macroscopic polarization mentioned earlier is $P(t) \propto \exp(-t/2T_1)$.[18] Hence, the macroscopic polarization decays at half the rate of the population decay. For simplicity, the factor of two between population and dephasing is included in all the time constants reported here and only population lifetimes are discussed (dephasing time constants are reported in Table SM2 for completeness). Different experimental parameters and checks required to ensure that accurate lifetimes are measured are discussed in the Methods section. The kinetic traces yielding lifetimes are obtained by taking lineouts at each feature's central frequencies.

The central energies of the four transitions discussed here are determined by fitting the static spectrum with lineshapes obtained by considering expected vibronic progressions, convoluted with the 50 meV spectral resolution and varying the values of the line broadening due to lifetime (cf. SM1 for details). Fig. 2b and 2c then show the kinetic fits at each feature's central frequencies reported at the positions of the arrows in Fig. 2a. In ICl, the decomposition of the static spectrum in Fig. 1a indicates that the features partially overlap, potentially leading to errors in measured lifetimes. This analysis is considered in SM2.

## Discussion

**Spectral assignment**

The states investigated are located near the iodine $N_{4,5}$ edge and correspond to $4d^{-1}6p$ core-excited states. Similar to other halogen containing diatomic molecules,[28] $4d$ core-levels of iodine are split into five different core-levels with three possible angular momentum projections ($L_z = 0$, 1 or 2) and two spin states ($S = \pm ½$). These projections represent different alignments and shapes of the hole wavefunction with respect to the I-Cl bond defining the z-axis. Experimentally, due to the finite spectrometer resolution of 50 meV, lineshape broadening of approximately 130 meV from vibrational effects, and a further 75 meV to 180 meV of lifetime broadening (depending on the states), only two types of hole alignment (angular momentum projections) are discernible, i.e. $L_z = 0, \pm 1$ with $d_{z^2}$, $d_{xz}$ and $d_{yz}$ character, and $L_z = \pm 2$ with $d_{xy}$ and $d_{x^2-y^2}$ character. Iodine core-level wavefunctions with contributions along the molecular axis (i.e. with $L_z = 0, \pm 1$) are shown in the orange box in Fig. 1c and correspond to core-MO $4d\sigma_{1/2}$, $4d\pi_{1/2}$ and $4d\pi_{3/2}$. Core-level MOs perpendicular to the molecular axis (i.e. with $L_z = \pm 2$) are shown in the blue box of Fig. 1c and correspond to $4d\delta_{3/2}$ and $4d\delta_{5/2}$.

Nominally, $6p$ Rydberg orbitals are assigned to $6p\sigma$ and $6p\pi$ states, following similar assignments for the $4d$ orbital. Due to the large radius of Rydberg orbitals, the $6p\sigma/6p\pi$ splitting is too small (< 50 meV) compared to the transition linewidths to be resolved. Both orbitals will thus be referred to as $6p$ without further distinction. Transitions to other Rydberg states, e.g. $6s$ or $7p$, are possible, but because of the relatively weak transition dipole moments to these states, transitions to $6p$ Rydberg states dominate the static absorption spectrum in the spectral region considered.[19]

**Molecular dynamics**

From the measured lifetimes, the timescales for electronic decay are much shorter than nuclear motion. The core-excited states discussed here are bound and the computed I-Cl stretch frequencies are between 420 cm$^{-1}$ to 430 cm$^{-1}$ depending on the state considered (cf. SM1 for details). Hence, the half vibrational period is approximately 40 fs and the nuclear displacement for a half vibration is 0.34 Å, as inferred from the potential energy curves shown in Fig S1a. Depending on the state considered, the timescale for the population lifetime gives a variation on the average internuclear distance (2.32 Å) of between 1% and 3.5% during one time constant of the electronic decay. This leads to the conclusion that the core-hole decays in ICl are an example of nearly pure electronic molecular dynamics.

Measured lifetimes show substantial dependence on the alignment of the core-hole MO relative to the molecular axis. As shown in Table 1, core-excited states aligned parallel to the covalent bond, i.e. with $L_z = 0, \pm 1$ and colored in orange in Fig. 1, are 1.9 and 1.6 times shorter-lived than the states aligned perpendicular to the molecular bond (with $L_z = \pm 2$ and colored in blue in Fig. 1). A similar effect was computed for the decay of van der Waals dimers[14,15] and points to a manifestation of a non-local effect on the iodine core-hole states. To confirm the presence of these effects in ICl we consider the decay channels open to the core-excited states.

Table 1: Iodine $4d^{-1}6p$ CESs lifetimes in ICl

| States | $4d\delta_{5/2}6p$ | $4d(\pi_{3/2};\sigma_{1/2})6p$ | $4d\delta_{3/2}6p$ | $4d\pi_{1/2}6p$ |
|---|---|---|---|---|
| Energy (eV) | 55.6 | 55.9 | 57.4 | 57.7 |
| Lifetime (fs) From ATAS measurement | 6.5 ± 0.6 | 3.5 ± 0.4 | 6.9 ± 0.6 | 4.3 ± 0.4 |

The $4d^{-1}6p$ excited state can decay by two types of pathways as represented in Fig. 3: participator channels (channel i in Fig. 3), where the electron in the 6p Rydberg orbital is ionized or moves to a lower-lying orbital, and spectator channels (channels ii and iii in Fig. 3) where only the valence and core-electrons are involved in the decay. Even though participator channels lead to lower energy products, they are known to be minor in atomic iodine[29] and xenon.[30] Spectator channels are favored in these atoms because of the stronger Coulombic interaction of the core-electron with valence electrons relative to the electron in the delocalized 6p Rydberg state. The situation in ICl is similar so the participator channels are not expected to contribute significantly to the core-excited state decay.

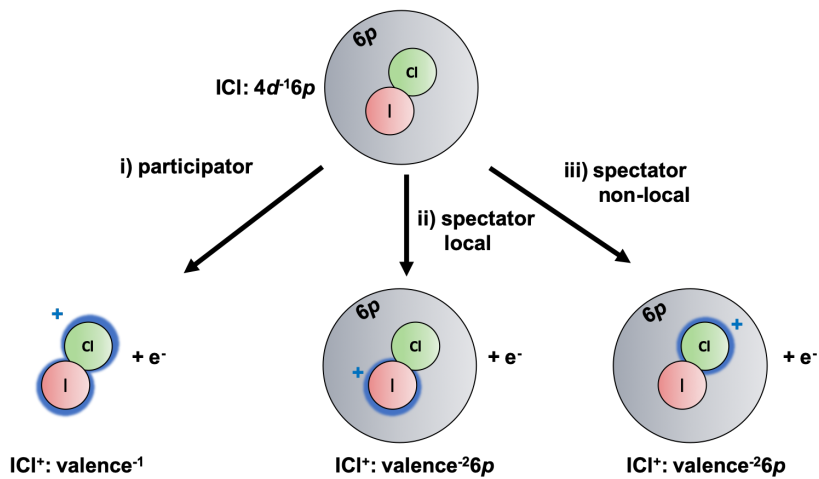

**Figure 3** Open decay channels for $4d^{-1}6p$ core-excited state. Channel i represents the participator channels where the electron in the Rydberg 6p orbital (represented in grey) is involved in the electronic decay. The local and non-local channels of the spectator decay lead to a partial charge on the iodine or chlorine atoms, respectively, as illustrated by the blue halos.

To understand the variation of lifetime with core-hole alignment, *ab initio* calculations of the core-hole partial linewidth were conducted using the Fano-ADC-Stieltjes method as detailed in the Methods section.[21,22] The calculation yields the partial linewidth associated with each open decay channel, which can be converted to lifetime using $\tau = \hbar/\Gamma$, where $\Gamma$ is the linewidth, and $\tau$ the lifetime. The method relies on a non-relativistic Hamiltonian, so effects such as spin-orbit interaction cannot be reproduced. The calculation of partial linewidths of core-excited states is difficult due to the complexity of the final state manifold. However, as core-hole relaxation is expected to be dominated by spectator decay, it is a reasonable approximation to consider the core ionized molecule rather than the $4d^{-1}6p$ state of the neutral. Therefore, we computed partial linewidths of ICl cations with a hole in the $4d$ orbital. In these conditions, the three cationic states considered are the $^2\Sigma$, $^2\Pi$ and $^2\Delta$ states, which correspond to the ionization limits of the $4d\sigma np$, $4d\pi np$ and $4d\delta np$ Rydberg series, respectively.

At the energy considered, 48 channels are open (cf. SM3). The decay channel showing the largest variation of partial linewidth with the type of initial core-hole is displayed in Fig. 4a. Given that the calculation considers core-excited cations as initial states, decay products are doubly charged with the two holes located in available valence MOs. For the channel considered in Fig. 4a, the two holes of this decay product are located in the same valence MO shown in the inset of Fig. 4a. For this final state, partial linewidths for the $^2\Sigma$, $^2\Pi$ and $^2\Delta$ core-holes are 14.1, 5.3 and 2.3 meV, respectively. The evolution of the linewidths with the core-hole types shows the decrease of the contribution of this final state to the core-hole decay. The main atomic orbital contributing to the empty MO shown is the Cl $3p$ (at 80%) so the contribution of non-local effects on the iodine core-hole decay is major for this channel. This highlights the role of non-local effects and core-orbital alignment on the core-hole decay.

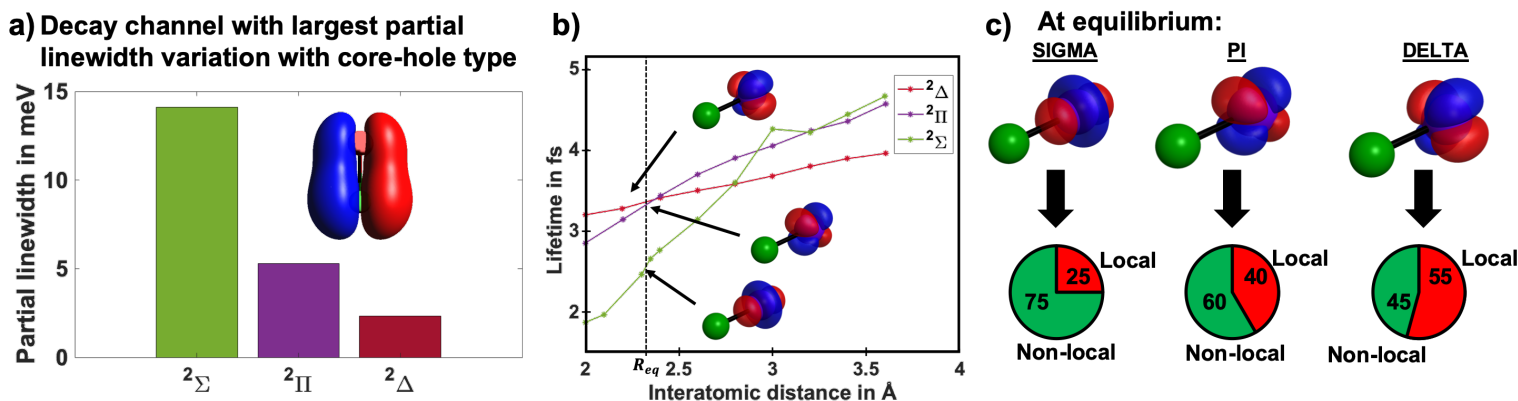

**Figure 4 a)** Partial linewidths of each of the core-hole types for the decay channel showing the largest partial linewidth variation. For this channel, the two valence holes of the decay product are located in the same MO shown in the inset. **b)** Calculated core-hole lifetime of the cation for the three types of core-hole orbitals at various internuclear distances. Conversion between linewidth and lifetime is obtain using $\tau = \hbar/\Gamma$, where $\Gamma$ is the linewidth, and $\tau$ the lifetime. The equilibrium distance (2.32 Å) is shown by the dotted line. **c)** Contribution of the local and non-local decay chanels at the equilibrium internuclear distance.

In order to compare experimental and computational results, all the decay channels have been considered and the MO of all the product and partial widths are shown in SM3. The total linewidths of each of the hole types are obtained by summing the partial linewidths of all open decay channels.

The lifetimes computed from the total widths are shown in Fig. 4b and are approximately twice as short as the experimental ones. This is due to a known bias in the computational method and consistent with other known cases.[31–33] Despite the approximations in the calculations, the relative simulated lifetimes capture the dependence on the core-hole alignment at various internuclear distances. At the equilibrium distance, ($R_{eq}$ = 2.32 Å),[34] the computed lifetime of the $^2\Sigma$ state is 1.25 times shorter than the lifetimes of $^2\Pi$ and $^2\Delta$. This can be compared to the ratio of the $4d(\pi_{3/2}; \sigma_{1/2})6p$ and $4d\delta_{5/2}6p$ CES lifetimes measured by ATAS of 1.9. The lifetimes computed for the cations in the $^2\Pi$ and $^2\Delta$ states at the equilibrium bond distance are similar despite the differences in the core-hole wavefunction alignments. This may reflect inaccuracies in the lifetime calculation due to the neglect of spin-orbit effects or consideration of the core-excited cation rather than the neutral species.

Computations consider the effect of the neighboring chlorine atom on the measured iodine core-hole lifetimes. In order to classify the valence delocalized orbitals into local and non-local channels, the final state empty MOs bearing the charges are projected onto atomic orbitals. Weights of the two types of channels are shown for each core-hole type at the equilibrium distance in Fig. 4c. While the $^2\Pi$ and $^2\Delta$ core-orbitals show similar contributions of local and non-local channels, the $^2\Sigma$ core-hole shows a major contribution of non-local effects.

## Conclusions

Molecular core-excited states dynamics of iodine monochloride (ICl) following $6p \leftarrow 4d$ core-to-Rydberg excitation are measured using ATAS. We show that this technique gives a direct access to lifetimes of the core-hole and allows one to follow the molecular decay in real time. Four molecular core-excited state lifetimes between 3.5 and 6.9 fs are reported depending on the core-hole character.

During the decay of the core-excited states with $4d^{-1}6p$ character the nuclei move by less than 3.5 % of their internuclear distance. Following on the ever-growing interest in charge migration[35,36] and photoionization time delays,[37] this investigation shows not only that the molecular dynamics of core states can be investigated by attosecond spectroscopy, but that their decays can be a nearly pure electronic decay process.

In this regime of molecular dynamics, core-excited state lifetimes depend on the core-hole orbital alignment with respect to the molecular axis. Core-hole orbitals with $4d\sigma_{1/2}$, $4d\pi_{1/2}$ and $4d\pi_{3/2}$ character show a reduction of their lifetimes compared to orbitals with $4d\delta_{3/2}$ and $4d\delta_{5/2}$ character. Calculations confirm this effect and attribute it to the larger contribution of non-local effects on the decay due to the presence of the nearby chlorine atoms for hole orbitals aligned along the molecular axis.

Molecular core-levels are often considered to be little influenced by the valence structure or molecular environments and are often replaced by pseudopotentials in electronic structure calculations in order to simplify relativistic effects.[38] Previous studies showed that core-level spectra can be non-degenerate due to ligand-field splitting.[13] This time-resolved study shows that their lifetimes are also greatly influenced by the molecular structure. This opens questions on the effect of lower molecular symmetry, ligand electronegativity, solvation environment etc. on core-excited state lifetimes. These topics relate to a wide range of fields such as chemistry and can now be addressed using attosecond spectroscopy.

## Methods:

**ATAS experiment:**
The laser setup was reported elsewhere[26]. Briefly, a carrier envelope phase stable Ti:Sapphire oscillator (Femtolaser, Rainbow) and multi-pass amplifier are used to produce carrier envelope stabilized pulses (1.8 mJ, 1 kHz, 25 fs at 780 nm). The laser pulses are spectrally broadened in a stretched hollow-core fiber of 2 m length and a 500 μm inner diameter (few-cycle Inc.) filled with Neon gas. Pulses are compressed using seven pairs of double angled chirped mirrors (Ultrafast Innovations, PC70) and a 3 mm thick ammonium dihydrogen phosphate crystal to correct for third order dispersion.[39] The compressed beam is separated by a 70/30 broadband beam splitter (LAYERTEC) and directed toward the high-order harmonic generation (HHG) cell and probe beam, respectively. Isolated attosecond pulses (IAPs) are generated by focusing the 3.7 fs long pulse (1.4 optical cycle) into a gas cell with flowing argon using a f = 500 mm concave mirror.[39] The driving field characterization was done using a commercially available dispersion scan, i.e. d-scan, (Sphere Ultrafast Photonics) and the results are shown in SM4. The isolated character of the attosecond pulse was established by confirming that the spectrum was continuous and showed a strong variation with the laser carrier envelope phase of the driving pulse, based on previous streaking measurements.[39]

The driving NIR field for the HHG is separated from the IAPs using a 200 nm thick aluminum filter supported on a mesh (Lebow). The IAP pulses are refocused toward the sample gas cell using a gold-coated toroidal mirror. The delayed NIR pulse is recombined with the IAP between the toroidal mirror and the target cell using an annular mirror. The delayed NIR pulse is focused using a silver coated f=1000 mm concave mirror. After the target cell, the NIR probe pulse is then removed using an aluminum filter similar that used to separate the HHG. An aberration-corrected concave grating (Hitachi, part number 001-0640) is used to disperse the light onto a CCD camera (Princeton Instrument, Pixis).[40] The experiment is conducted using the diffraction grating in second order and the spectral resolution was determined to be 50 meV at 65 eV by fitting of the core-level transitions in xenon.

The ICl sample was purchased from Sigma Aldrich and used without further purification. Adequate sample density was achieved by heating the sample container and gas lines to 40 °C using heat tapes. Transient absorption spectra were obtained by collecting the IAP spectrum with ($I_{on}$) and without the delayed NIR beam ($I_{off}$) (modulated using a mechanical shutter) and computed using $\Delta A = -\log(\frac{I_{on}}{I_{off}})$.

**Core-hole lifetime measurement**
To reduce systematic errors in lifetime measurements due to low spectral resolution, inducing an early cancellation of the depletion feature by nearby positive features, static and time-resolved data were collected using the second order of the spectrometer diffraction grating reaching a spectral resolution of 50 meV. Power dependences and discussion of the spectral resolution are detailed in the SM5 to verify that the measurements accurately report core-hole lifetimes. To further confirm the lifetime measurement, the lifetime of core-excited xenon following excitation of its $6p \leftarrow 4d$ transition was measured using ATAS under the same laser power, pressure and spectral resolution conditions as those used in the ICl experiments. A lifetime of $5.9 \pm 0.7$ fs is measured in these conditions, in good agreement with a previous estimate from linewidth measurements, which indicated a lifetime of $6.2 \pm 0.2$ fs.[41] (cf. SM6 for spectrum and kinetic traces).

**Fano-ADC-Stieltjes method**

Ab initio calculations of the core-hole lifetime were conducted using the Fano-algebraic diagrammatic construction (ADC)-Stieltjes method[21,22] In the Fano-ADC-Stieltjes method, natural linewidths (Γ) are obtained by separately constructing the continuum composed of the decaying state and the leaving electron ($\chi_{\beta,\varepsilon}$), the bound initial state (Φ), and the coupling between the two. The width is given by the golden rule-like expression, where the coupling moments are summed over all open decay channels.

$$\Gamma = 2\pi \sum_{\beta} |\langle \Phi | \hat{H} | \chi_{\beta,\varepsilon} \rangle|^2$$

The width is then converted to lifetime following the uncertainty principle ($\tau = \hbar/\Gamma$).

As discussed in the main text, the decay of core-excited states considered mainly occurs via spectator mechanisms where the electron in the 6$p$ Rydberg orbital does not participate. To simplify the calculation, core-ionized cations were, therefore, considered instead of core excited ones. Moreover, the non-relativistic Hamiltonian has been used to construct the initial and final states and, thus, the spin-orbit interactions have been neglected. In these approximations, three core ionized states are considered: the $^2\Sigma$, $^2\Pi$ and $^2\Delta$, corresponding to the ionization limits of the $4d\sigma np$, $4d\pi np$ and $4d\delta np$ Rydberg series, respectively.

## Acknowledgements:

The authors acknowledge Lorenz Cederbaum and Romain Geneaux for valuable discussions. H.J.B.M., A.P.F., D.M.N. and S.R.L. acknowledge the Director, Office of Science, Office of Basic Energy Sciences through the Atomic, Molecular, and Optical Sciences Program of the Division of Chemical Sciences, Geosciences, and Biosciences of the U.S. Department of Energy at Lawrence Berkeley National Laboratory under contract no. DE-AC02-05CH11231. A.P.F. acknowledges funding from the NSF Graduate Research Fellowship Program. Y.K., D.M.N., S.R.L. and A.I.K. acknowledge support from the U.S. Army Research Office (ARO) (No. W911NF-14-1-0383). Y.K., and S.R.L. acknowledge support from the National Science Foundation (NSF) (CHE-1660417) for absorption spectra calculations Y.K. also acknowledges financial support from the Funai Overseas Scholarship. A.G, K.G. and A.I.K. acknowledge the support by the European Research Council (ERC) under the Advanced Investigator Grant No. 692657.

## Author Contributions:

H.J.B.M., A.P.F., D.M.N. and S.R.L. designed the experiment. H.J.B.M. and A.P.F. performed the data collection. A.G, K.G. and A.I.K. performed the core-hole lifetime calculations and Y.K. performed the absorption spectra calculations. H.J.B.M., D.M.N. and S.R.L. wrote the manuscript.

## Competing Interests statement:

The authors declare no competing interests.

## References:

(1) D. Chattarji. *The Theory of Auger Transitions*; Academic Press Inc, Ed.; New York, 1976.
(2) Föhlisch, A.; Feulner, P.; Hennies, F.; Fink, A.; Menzel, D.; Sanchez-Portal, D.; Echenique, P. M.; Wurth, W. Direct Observation of Electron Dynamics in the Attosecond Domain. *Nature* **2005**, *436* (7049), 373–376. https://doi.org/10.1038/nature03833.
(3) Püttner, R.; Marchenko, T.; Guillemin, R.; Journel, L.; Goldsztejn, G.; Céolin, D.; Takahashi, O.; Ueda, K.; Lago, A. F.; Piancastelli, M. N.; et al. Si 1s -1 , 2s -1 and 2p -1 Lifetime Broadening of SiX 4 (X = F, Cl, Br, CH 3 ) Molecules: SiF 4 Anomalous Behaviour Reassessed. *Phys. Chem. Chem. Phys.* **2019**, *21*, 8827–8836. https://doi.org/10.1039/c8cp07369d.
(4) Ramasesha, K.; Leone, S. R.; Neumark, D. M. Real-Time Probing of Electron Dynamics Using Attosecond Time-Resolved Spectroscopy. *Annu. Rev. Phys. Chem.* **2016**, *67*, 41–63. https://doi.org/10.1146/annurev-physchem-040215-112025.

(5)  Uiberacker, M.; Uphues, T.; Schultze, M.; Verhoef, A. J.; Yakovlev, V.; Kling, M. F.; Rauschenberger, J.; Kabachnik, N. M.; Schröder, H.; Lezius, M.; et al. Attosecond Real-Time Observation of Electron Tunnelling in Atoms. *Nature* **2007**, *446*, 627–632. https://doi.org/10.1038/nature05648.

(6)  Hütten, K.; Mittermair, M.; Stock, S. O.; Beerwerth, R.; Shirvanyan, V.; Riemensberger, J.; Duensing, A.; Heider, R.; Wagner, M. S.; Guggenmos, A.; et al. Ultrafast Quantum Control of Ionization Dynamics in Krypton. *Nat. Commun.* **2018**, *9*, 719. https://doi.org/10.1038/s41467-018-03122-1.

(7)  Drescher, L.; Reitsma, G.; Witting, T.; Patchkovskii, S.; Mikosch, J.; Vrakking, M. J. J. State-Resolved Probing of Attosecond Timescale Molecular Dipoles. *J. Phys. Chem. Lett.* **2019**, *10* (2), 265–269. https://doi.org/10.1021/acs.jpclett.8b02878.

(8)  Cederbaum, L. S.; Zobeley, J.; Tarantelli, F. Giant Intermolecular Decay and Fragmentation of Clusters. *Phys. Rev. Lett.* **1997**, *79*, 4778. https://doi.org/10.1103/PhysRevLett.79.4778.

(9)  Zobeley, J.; Santra, R.; Cederbaum, L. S. Electronic Decay in Weakly Bound Heteroclusters: Energy Transfer versus Electron Transfer. *J. Chem. Phys.* **2001**, *115* (11), 5076–5088. https://doi.org/10.1063/1.1395555.

(10) Schnorr, K.; Senftleben, A.; Kurka, M.; Rudenko, A.; Foucar, L.; Schmid, G.; Broska, A.; Pfeifer, T.; Meyer, K.; Anielski, D.; et al. Time-Resolved Measurement of Interatomic Coulombic Decay in Ne2. *Phys. Rev. Lett.* **2013**, *111*, 093402. https://doi.org/10.1103/PhysRevLett.111.093402.

(11) Takanashi, T.; Golubev, N. V.; Callegari, C.; Fukuzawa, H.; Motomura, K.; Iablonskyi, D.; Kumagai, Y.; Mondal, S.; Tachibana, T.; Nagaya, K.; et al. Time-Resolved Measurement of Interatomic Coulombic Decay Induced by Two-Photon Double Excitation of Ne2. *Phys. Rev. Lett.* **2017**, *118*, 033202. https://doi.org/10.1103/PhysRevLett.118.033202.

(12) Mizuno, T.; Cörlin, P.; Miteva, T.; Gokhberg, K.; Kuleff, A.; Cederbaum, L. S.; Pfeifer, T.; Fischer, A.; Moshammer, R. Time-Resolved Observation of Interatomic Excitation-Energy Transfer in Argon Dimers. *J. Chem. Phys.* **2017**, *146*, 104305. https://doi.org/10.1063/1.4978233.

(13) Cutler, J. N.; Bancroft, G. M.; Tan, K. H. Ligand-Field Splittings and Core-Level Linewidths in I 4d Photoelectron Spectra of Iodine Molecules. *J. Chem. Phys.* **1992**, *97*, 7932. https://doi.org/10.1063/1.463468.

(14) Kryzhevoi, N. V; Averbukh, V.; Cederbaum, L. S. High Activity of Helium Droplets Following Ionization of Systems inside Those Droplets. *Phys. Rev. B* **2007**, *76* (9), 94513. https://doi.org/10.1103/PhysRevB.76.094513.

(15) Gokhberg, K.; Kopelke, S.; Kryzhevoi, N. V.; Kolorenč, P.; Cederbaum, L. S. Dependence of Interatomic Decay Widths on the Symmetry of the Decaying State: Analytical Expressions and Ab Initio Results. *Phys. Rev. A - At. Mol. Opt. Phys.* **2010**, *81*, 013417. https://doi.org/10.1103/PhysRevA.81.013417.

(16) Beck, A. R.; Bernhardt, B.; Warrick, E. R.; Wu, M.; Chen, S.; Gaarde, M. B.; Schafer, K. J.; Neumark, D. M.; Leone, S. R. Attosecond Transient Absorption Probing of Electronic Superpositions of Bound States in Neon: Detection of Quantum Beats. *New J. Phys.* **2014**, *16* (11), 113016. https://doi.org/10.1088/1367-2630/16/11/113016.

(17) Wang, H.; Chini, M.; Chen, S.; Zhang, C.-H.; He, F.; Cheng, Y.; Wu, Y.; Thumm, U.; Chang, Z. Attosecond Time-Resolved Autoionization of Argon. *Phys. Rev. Lett.* **2010**, *105*

(14), 143002. https://doi.org/10.1103/PhysRevLett.105.143002.
(18) Bernhardt, B.; Beck, A. R.; Li, X.; Warrick, E. R.; Bell, M. J.; Haxton, D. J.; McCurdy, C. W.; Neumark, D. M.; Leone, S. R. High-Spectral-Resolution Attosecond Absorption Spectroscopy of Autoionization in Xenon. *Phys. Rev. A - At. Mol. Opt. Phys.* **2014**, *89*, 023408. https://doi.org/10.1103/PhysRevA.89.023408.
(19) Brion, C. E.; Dyck, M.; Cooper, G. Absolute Photoabsorption Cross-Sections (Oscillator Strengths) for Valence and Inner Shell Excitations in Hydrogen Chloride, Hydrogen Bromide and Hydrogen Iodide. *J. Electron Spectros. Relat. Phenomena* **2005**, *144–147*, 127–130. https://doi.org/10.1016/j.elspec.2005.01.010.
(20) Olney, T. N.; Cooper, G.; Brion, C. E. Quantitative Studies of the Photoabsorption (4.5-488 EV) and Photoionization (9-59.5 EV) of Methyl Iodide Using Dipole Electron Impact Techniques. *Chem. Phys.* **1998**, *232* (1–2), 211–237. https://doi.org/10.1016/S0301-0104(97)00368-6.
(21) Santra, R.; Cederbaum, L. S. Non-Hermitian Electronic Theory and Applications to Clusters. *Phys. Rep.* **2002**, *368* (1), 1–117. https://doi.org/10.1016/S0370-1573(02)00143-6.
(22) Averbukh, V.; Cederbaum, L. S. Ab Initio Calculation of Interatomic Decay Rates by a Combination of the Fano Ansatz, Green's-Function Methods, and the Stieltjes Imaging Technique. *J. Chem. Phys.* **2005**, *123*, 204107. https://doi.org/10.1063/1.2126976.
(23) Marroux, H. J. B.; Fidler, A. P.; Neumark, D. M.; Leone, S. R. Multidimensional Spectroscopy with Attosecond Extreme Ultraviolet and Shaped Near-Infrared Pulses. *Sci. Adv.* **2018**, *4* (9).
(24) Wu, M.; Chen, S.; Camp, S.; Schafer, K. J.; Gaarde, M. B. Theory of Strong-Field Attosecond Transient Absorption. *J. Phys. B At. Mol. Opt. Phys.* **2016**, *49* (6), 062003. https://doi.org/10.1088/0953-4075/49/6/062003.
(25) Kaldun, A.; Blättermann, A.; Stooß, V.; Donsa, S.; Wei, H.; Pazourek, R.; Nagele, S.; Ott, C.; Lin, C. D.; Burgdörfer, J.; et al. Observing the Ultrafast Buildup of a Fano Resonance in the Time Domain. *Science (80-. ).* **2016**, *354* (6313), 738–741. https://doi.org/10.1126/science.aah6972.
(26) Cao, W.; Warrick, E. R.; Neumark, D. M.; Leone, S. R. Attosecond Transient Absorption of Argon Atoms in the Vacuum Ultraviolet Region: Line Energy Shifts versus Coherent Population Transfer. *New J. Phys.* **2016**, *18*, 013041. https://doi.org/10.1088/1367-2630/18/1/013041.
(27) Ott, C.; Kaldun, A.; Raith, P.; Meyer, K.; Laux, M.; Evers, J.; Keitel, C. H.; Greene, C. H.; Pfeifer, T. Lorentz Meets Fano in Spectral Line Shapes: A Universal Phase and Its Laser Control. *Science (80-. ).* **2013**, *340* (6133), 716–720. https://doi.org/10.1126/science.1234407.
(28) Johnson, J.; Cutler, J. N.; Bancroft, G. M.; Hu, Y. F.; Tan, K. H. High-Resolution Photoabsorption and Photoelectron Spectra of Bromine-Containing Molecules at the Br 3d Edge: The Importance of Ligand Field Splitting. *J. Phys. B At. Mol. Opt. Phys.* **1997**, *30* (21), 4899. https://doi.org/10.1088/0953-4075/30/21/024.
(29) Nahon, L.; Morin, P. Experimental Study of Rydberg States Excited from the d Shell of Atomic Bromine and Iodine. *Phys. Rev. A* **1992**, *45*, 2887. https://doi.org/10.1103/PhysRevA.45.2887.
(30) Southworth, S.; Becker, U.; Truesdale, C. M.; Kobrin, P. H.; Lindle, D. W.; Owaki, S.; Shirley, D. A. Electron-Spectroscopy Study of Inner-Shell Photoexcitation and Ionization of Xe. *Phys. Rev. A* **1983**, *28* (1), 261. https://doi.org/10.1103/PhysRevA.28.261.

(31) Stumpf, V.; Brunken, C.; Gokhberg, K. Impact of Metal Ion's Charge on the Interatomic Coulombic Decay Widths in Microsolvated Clusters. *J. Chem. Phys.* **2016**, *145*, 104306. https://doi.org/10.1063/1.4962353.

(32) Trinter, F.; Williams, J. B.; Weller, M.; Waitz, M.; Pitzer, M.; Voigtsberger, J.; Schober, C.; Kastirke, G.; Müller, C.; Goihl, C.; et al. Evolution of Interatomic Coulombic Decay in the Time Domain. *Phys. Rev. Lett.* **2013**, *111*, 093401. https://doi.org/10.1103/PhysRevLett.111.093401.

(33) Ouchi, T.; Sakai, K.; Fukuzawa, H.; Higuchi, I.; Demekhin, P. V.; Chiang, Y. C.; Stoychev, S. D.; Kuleff, A. I.; Mazza, T.; Schöffler, M.; et al. Interatomic Coulombic Decay Following Ne 1s Auger Decay in NeAr. *Phys. Rev. A - At. Mol. Opt. Phys.* **2011**, *83*, 053415. https://doi.org/10.1103/PhysRevA.83.053415.

(34) Herbst, E.; Steinmetz, W. Dipole Moment of ICl. *J. Chem. Phys.* **1972**, *56* (11), 5342–5346. https://doi.org/10.1063/1.1677044.

(35) Calegari, F.; Ayuso, D.; Trabattoni, A.; Belshaw, L.; De Camillis, S.; Anumula, S.; Frassetto, F.; Poletto, L.; Palacios, A.; Decleva, P.; et al. Ultrafast Electron Dynamics in Phenylalanine Initiated by Attosecond Pulses. *Science (80-. ).* **2014**, *346* (6207), 336–339. https://doi.org/10.1126/science.1254061.

(36) Kuleff, A. I.; Cederbaum, L. S. Ultrafast Correlation-Driven Electron Dynamics. *J. Phys. B At. Mol. Opt. Phys.* **2014**, *47* (12), 124002. https://doi.org/10.1088/0953-4075/47/12/124002.

(37) Beaulieu, S.; Comby, A.; Clergerie, A.; Caillat, J.; Descamps, D.; Dudovich, N.; Fabre, B.; Géneaux, R.; Légaré, F.; Petit, S.; et al. Attosecond-Resolved Photoionization of Chiral Molecules. *Science (80-. ).* **2017**, *358* (6368), 1288–1294. https://doi.org/10.1126/science.aao5624.

(38) Dolg, M. Effective Core Potentials. *Mod. Methods Algorithms Quantum Chem.* **2000**, *1*, 479–508. https://doi.org/10.1002/qua.560530307.

(39) Timmers, H.; Kobayashi, Y.; Chang, K. F.; Reduzzi, M.; Neumark, D. M.; Leone, S. R. Generating High-Contrast, near Single-Cycle Waveforms with Third-Order Dispersion Compensation. *Opt. Lett.* **2017**, *42* (4), 811–814. https://doi.org/10.1364/OL.42.000811.

(40) Wang, X.; Chini, M.; Cheng, Y.; Wu, Y.; Chang, Z. In Situ Calibration of an Extreme Ultraviolet Spectrometer for Attosecond Transient Absorption Experiments. *Appl. Opt.* **2013**, *52* (3), 323–329.

(41) Penent, F.; Palaudoux, J.; Lablanquie, P.; Andric, L.; Feifel, R.; Eland, J. H. D. Multielectron Spectroscopy: The Xenon 4d Hole Double Auger Decay. *Phys. Rev. Lett.* **2005**, *95* (8), 83002. https://doi.org/10.1103/PhysRevLett.95.083002.